# A model for retention on short, intermediate and long time-scale in ferroelectric thin films

X.J. Lou*

*Department of Materials Science and Engineering, National University of Singapore, 117574, Singapore*

**Abstract:**

We developed a model with *no* adjustable parameter for retention loss at short and long time scale in ferroelectric thin-film capacitors. We found that the predictions of this model are in good agreement with the experimental observations in the literature. In particular, it explains why a power-law function shows better fitting than a linear-log relation on a short time scale ($10^{-7}$ s to 1 s) and why a stretched exponential relation gives more precise description than a linear-log plot on a long time scale (>100 s), as reported by many researchers in the past. More severe retention losses at higher temperatures and in thinner films have also been correctly predicted by the present theory.

Retention, the ability of maintaining a poled polarization state with time, has long been one of the most important reliability issues hindering the commercialization of ferroelectric memories, along with imprint and fatigue [1, 2]. It has been generally observed that a fast polarization decay or relaxation occurs on a short-time scale (e.g. $t$<1 s), followed by a relatively slow decay and then a long tail on an intermediate and long-time scale (e.g. $t$>1 s) in linear-log plot [3, 4]. Although the problem of retention has been studied for a few decades, the mechanism dominating the retention loss at both short-time and long-time scale remains largely unsolved. It has been generally accepted that the depolarization field, which is induced by imperfect screening at the electrode-film interface, is the main reason for short-time retention loss [5]. But how the depolarization field induces polarization loss via backswitching on a short-time scale and to what time extent the concept of depolarization field still applies remain unknown. In addition, different research groups have focused on the problem of retention loss at different time-scales and have fitted their data using different empirical functions, such as linear-log [3, 6, 7], power-law [8-10], and a stretched exponential relation [6, 7, 11, 12]. From the above discussion, one can see that the microscopic mechanism dominating the problem of retention loss at both short and long time scale in ferroelectric thin films is still unclear and very controversial. The standard way to obtain the retention properties of a ferroelectric capacitor in the literature is to empirically extrapolate the data measured over a shorter period of time (up to $10^6$ s) to 10 years (the required shelf-life for a ferroelectric memory cell) without any physical justifications.

* Correspondence email: mselx@nus.edu.sg

In this paper, we develop a polarization retention model applied for short ($10^{-7}$ s), intermediate (~1 s) and long time-scales (>10 years) in ferroelectric capacitors with *no* adjustable parameter. The predictions of this model and its consistency with the results in the literature are also discussed.

It is well known that even though a ferroelectric capacitor is fully poled by a saturating field the depolarization field *unavoidably* arises to some extent due to the incomplete screening of polarization at the electrode-film interface after the external field is removed [3]. Furthermore, the depolarization field seen by a poled Pt/PZT/Pt thin film [PZT denotes $Pb(Zr,Ti)O_3$] has been shown to be [13]:

$$E_{dep}(t) = \frac{d_i P(t)}{d\varepsilon_i \varepsilon_0} \qquad (1)$$

where $d_i$ is the thickness of the interface layer. Inserting $P$=30 μC/cm$^2$, $\varepsilon_i/d_i$=20 nm$^{-1}$ (we assume $\varepsilon_i$=40 and $d_i$=2 nm [14]) and $d$=200 nm for a standard ferroelectric Pt/PZT/Pt capacitor, we have $E_{dep}$=85 kV/cm, comparable to the threshold field for switching in PZT thin films. Therefore, backswitching due to the polarization instability caused by depolarization field is expected to induce retention loss in ferroelectric thin films, as already discussed in detail in the literature.

Note that depolarization field may appear not only for fully poled Pt/PZT/Pt structure with interface layers, but also for $M_{oxide}$/PZT/ $M_{oxide}$ and Pt/SBT($SrBi_2Ta_2O_9$)/Pt, which are believed to contain no "dead layer" [15]. As argued by Benedetto *et al*., depolarization field always exists to some extent in insulating thin-film ferroelectrics due to a passive layer in ferroelectric-electrode interface (i.e., the case discussed above), a finite separation between polarization charge and electrode screening charge, a gradient in polarization near electrode, or distribution of the electrode screening charge over a small but finite distance [3]. Therefore, in general, the depolarization field due to poor screening of polarization bound charges can be written as:

$$E_{dep}(t) = \beta \frac{P(t)}{\varepsilon_f \varepsilon_0} \qquad (2)$$

where $\beta$ is the depolarization factor and is generally less than 1, $\varepsilon_f$ is the dielectric constant of the ferroelectric.

Then we assume that backswitching that causes polarization loss with time is nothing more than polarization switching driven purely by a time-dependent depolarization field $E_{dep}(t)$. Backswitching and the resultant formation of opposite domains manifest themselves as fast polarization relaxation on a short time scale (<1 s) and "retention loss" on a long time scale (>>1 s).

Let us imagine that the total surface area of a poled ferroelectric capacitor is divided *uniformly* into $M_0$ ($M_0$>>1, an integer) parts. Then we assume that the capacitor backswitches in part-by-part or region-by-region manner due to the hindering effect of grain boundaries [16], defect planes/dislocations and/or 90 ° domain walls [17], consistent with what have been observed in ferroelectric thin films [16-18]. In other words, it occurs via nucleation of opposite domain (the key and time-dominating process), followed by much quicker forward and sideways growth of domain until it reaches the boundary of this part. That is to say, domain wall motion is only allowed within each part/region; domain wall motion crossing the boundaries between one part/region and another is forbidden in our current model. Since both $P(t)$ and $E_{dep}(t)$ relax with time (see Eq. (1) and Eq (2)), we have to include a *feedback* mechanism for $E_{dep}(t)$ in the following derivation. In other words, we will consider a feedback loop of the depolarization field corrected by the updated retained polarization at each time point, where one more part has just switched.

Let us begin with a fully poled ferroelectric capacitor with retained polarization $P_{M_0}$ at $t=t_0=0\sim10^{-13}$ s (the period of soft mode) when the external field is *just* removed. Then we assume that $1/\zeta(t_N)$ ($\zeta>>1$) is the backswitching probability for one of the retained parts after $t_c$ from the time point where the $N$th part has just backswitches. $t_N$ is defined as the time interval that the $N$th part takes to backswitches. $t_c$ is a characteristic time and can be chosen *arbitrarily* as long as it ensures $1/\zeta(t_N)<<1$ for any $t_N$ (it disappears in the following derivation as we will see later). Therefore, the probability that one part will retain its polarization after $t_c$ from time $t_0$ is $[1-1/\zeta(t_0)]$. The probability that this part will survive from backswitching after $t_1$ ($t_1>>t_c$) from time $t_0$ is $\left(1-\frac{1}{\zeta(t_0)}\right)^{t_1/t_c}$. According to the definition of $t_1$, the time interval that the *first* part takes to switch, the total number of the parts that survive from backswitching after $t_1$ is:

$$M_0-1=M_0\left(1-\frac{1}{\zeta(t_0)}\right)^{t_1/t_c} \quad (3)$$

which can be reformed to:

$$\frac{M_0-1}{M_0}=\left(1-\frac{1}{\zeta(t_0)}\right)^{t_1/t_c} \quad (4)$$

Taking natural logarithm on both sides of Eq (4), we have:

$$\ln \frac{M_0 - 1}{M_0} = \frac{t_1}{t_c} \ln\left(1 - \frac{1}{\zeta(t_0)}\right) = \frac{t_1}{t_c}\left[-\frac{1}{\zeta(t_0)} - \frac{1}{2}\left(\frac{1}{\zeta^2(t_0)}\right) + \cdots\right] \quad (5)$$

Because $\zeta(t_0) \gg 1$, all the higher-order terms can be neglected. So we have:

$$\frac{M_0 - 1}{M_0} = \exp\left(-\frac{t_1}{t_c \zeta(t_0)}\right) \quad (6)$$

Let us now consider Merz's law [19], which has been used to fit the switching data in both bulk and thin-film ferroelectrics [19-22]:

$$t_{sw} = t_\infty \cdot \exp\left(+\frac{\alpha}{E}\right) \text{ or } \frac{1}{t_{sw}} = \frac{1}{t_\infty} \cdot \exp\left(-\frac{\alpha}{E}\right) \quad (7)$$

where $t_{sw}$ and $t_\infty$ is the switching time for $E$ and an infinite field, respectively. $\alpha$ is the activation field for switching, and it depends on temperature $T$ and film thickness $d$ [19, 20]. $\alpha$ varies from 200 kV/cm to 800 kV/cm for ferroelectric thin films in the literature [1, 21-23]. Recalling the meaning of $1/\zeta(t_0)$ defined above, we have:

$$\frac{1}{\zeta(t_0)} = \frac{t_c}{t_{sw}(t_0)} = \frac{t_c}{t_\infty} \cdot \exp\left(-\frac{\alpha}{E_{dep}(t_0)}\right) \quad (8)$$

$t_\infty$ can be estimated both experimentally and theoretically for a specific sample. The fastest $t_{sw}$ measured so far is 220 ps from a circuit with a *RC* constant value of about 45 ps (on a 4.5 μm x 5.4 μm electrode) [23], in which $t_\infty$ should be ~100 ps. For a standard measurement circuit, however, $t_{sw}$ is around ~100 ns with $t_\infty$ estimated to be 13 ns by Scott *et al.* in submicron PZT thin films [20], which is comparable with Merz's results which show $t_\infty$ decreases with $d$, and $t_\infty$=10 ns in a 20 μm-thick BTO ($BaTiO_3$) single-crystal [19]. Theoretically, we can simply regard $t_\infty \sim t_{pg} = d/v$, $t_{pg}$ is the propagating time of a needlelike domain [1]. Assuming $d$=200 nm and $v$=2000 m/s, the sound velocity, we have $t_\infty \sim t_{pg}$=100 ps. Therefore, for a submicron thin-film ferroelectric capacitor $t_\infty$ varies from 100 ps to 10 ns depending on the parameters of a specific system; and $t_\infty$ in Eq(7) and Eq (8) above should be (much) smaller because it is the value for one part of the whole capacitor. For simplicity, we use $t_\infty$ =1 ns in this work. Note that due to the limitation of experimental instrument and setups, switching time less than 100 ps or polarization relaxation data collected at shorter than 1 ns pulse delay time are hard to achieve.

Inserting Eq (8) and Eq (1) into Eq (6), we have [for the ease of mathematical estimation, we will use Eq (1) rather than Eq (2) in the following derivation. However, readers would find it rather straightforward to replace Eq (1) by the general form Eq (2) or any other specific form of $E_{dep}$]:

$$\frac{M_0 - 1}{M_0} = \exp\left( -\frac{t_1}{t_\infty} \cdot \exp\left( -\frac{\alpha d \varepsilon_i \varepsilon_0}{d_i P_{M_0}} \right) \right) \qquad (9)$$

Similarly, we get a series of *feedback* equations:

$$\frac{M_0 - 2}{M_0 - 1} = \exp\left( -\frac{t_2}{t_\infty} \cdot \exp\left( -\frac{\alpha d \varepsilon_i \varepsilon_0}{d_i P_{M_0 - 1}} \right) \right)$$

$$\vdots$$

$$\frac{M_0 - N - 1}{M_0 - N} = \exp\left( -\frac{t_{N+1}}{t_\infty} \cdot \exp\left( -\frac{\alpha d \varepsilon_i \varepsilon_0}{d_i P_{M_0 - N}} \right) \right) \qquad (10)$$

$$\vdots$$

$$\frac{\frac{M_0}{2}}{\frac{M_0}{2} + 1} = \exp\left( -\frac{t_{M_0/2}}{t_\infty} \cdot \exp\left( -\frac{\alpha d \varepsilon_i \varepsilon_0}{d_i P_{\frac{M_0}{2}+1}} \right) \right)$$

where $N = \left(0, 1, \cdots, \frac{M_0}{2} - 1\right)$ and $P_{M_0 - N} = \frac{M_0 - 2N}{M_0} P_{M_0}$. Note that Eq (10) contains *no adjustable parameter*. All the parameters or physical quantities involved are experimentally measurable.

Let us make a few remarks about this model and make some predictions using it:

(1) The ($Y$, $X$) profiles of $\left(\frac{P_{M_0-N}}{P_{M_0}} = \frac{M_0 - 2N}{M_0}, t = \sum_{i=1}^{N} t_i\right)$ for $M_0$=100, 500 and 1000 have been plotted in Fig 1, where we assume $\alpha$=500 kV/cm [1, 21-23], $P_{M_0}$ =30 μC/cm$^2$, $d$=200 nm, $\varepsilon_i/d_i$=20 nm$^{-1}$ and $t_\infty$ =1 ns as justified above. Data points for 10 years (3.15x10$^8$ s) and the age of the Earth (1.43x10$^{17}$ s) have also been labeled. One can see that there is *no* analytical equation that can describe the full curve; each data point has to be constructed using Eq (10) to calculate $t_N$ and using $t = \sum_{i=1}^{N} t_i$ to get the total time. We also see that these three curves essentially overlap; the curves are stabilized when $M_0 \to \infty$ (e.g., the deviation between the curves of $M_0$=500 and 1000 is negligible). From Eq (10), we see that the effect of the change in $t_\infty$ is just to shift the whole curve slightly along the $X$ axis without changing its profile.

(2) From Fig 1, one can see that in a linear-log plot the normalized polarization shows a small plateau around 1 ns, followed by a dramatic decrease until 1 ms and then a slow decay and a long tail up to time infinity (Note that the experimentally achievable delay time is normally longer than 100 ns). This feature is consistent with the results in the literature [3, 4, 10]. If we assume that the retained polarization should be larger than 4 μC/cm$^2$ in order to provide enough switched charge for READ operation, then the model capacitor studied here will fail at 3x10$^7$ s (i.e. ~1 year). However, if the requirement is set to be 3 μC/cm$^2$, then the model capacitor can be still reliably detected even after 1 million years (3.9x10$^{13}$ s) due to the very slow decay in long time-scale. It can also be seen that the profile indeed shows a linear-log relationship between $P(t)$ and $t$ for some range of the curve (e.g. from 10$^{-9}$ s to 10$^{-4}$ s, or any section of the curve shorter than three decades), but not for the full range, which is also in agreement with the published data [3, 6, 7]. The power-law plot and the stretched exponential plot of the curve with $M_0$=1000 have been shown in Fig 2. This figure accounts for the observations of Jo and Kim *et al*. [9, 10], who showed that their data in a range of $10^{-6}$ s to 1 s follow a power-law $\left[P(t)\sim t^{-n}\right]$ behavior rather than a linear-log one. After fitting the curve part of interest, we get $n$=0.07 (see Fig 3, note that Fig 1 indicates a linear-log fitting for this time range is very poor), in excellent agreement with the data of Kang *et al*., who *also* got $n$=0.07 for their virgin PZT films [8], and in good agreement with the results of Jo *et al*. [10], who showed that a thickness-dependent $n$ value with $n$=0.12 for the 30 nm-thick BTO sample (note $n$ decreases as $d$ increases, see Fig 2c in Ref [10]). Fig 2 (in comparison with Fig 1) also explains the results of Hong and Jo *et al*., who found that the retention loss in long-time scale in some cases fits a linear-log dependence, however, in most cases it seems to fit a stretched exponential function $\left[P(t)=P_0\exp\left(-c\cdot t^m\right)\right]$ better [7]. After fitting the relevant part of our curve, we get $m$=0.0176 (see Fig 3), one order of magnitude lower than those observed by Hong *et al*., who got $m$=0.145, and 0.175 for top-electrode free PZT/$LaNiO_3$ and PZT/Pt, respectively [7]. That is reasonable because the samples they studied have no top electrodes and therefore should show much larger $E_{dep}$ and thus higher decay rates (i.e. higher $m$ values). Note that the *feedback* nature of our model requiring some pre-knowledge about $t_\infty$, $\alpha$, $P_{M_0}$ and $\varepsilon_i/d_i$ for a specific sample does not allow us to give further quantitative description about those relaxation data and/or fit them using software like Origin.

(3) From Eq (10), we obtain:

$$t_N=\frac{-t_\infty\cdot\ln\dfrac{M_0-N}{M_0-N+1}}{\exp\left(-\dfrac{\alpha(T,d)d\varepsilon_i\varepsilon_0}{d_iP_{M_0-N+1}\left(T,E,d\right)}\right)}\sim\exp\left(\frac{\alpha(T,d)d\varepsilon_i\varepsilon_0}{d_iP_{M_0-N+1}\left(T,E,d\right)}\right) \qquad \text{Eq (11)}$$

Eq (11) indicates that $t_N$ (and therefore $t = \sum_{i=1}^{N} t_i$ ) increases *exponentially* with $\frac{\alpha(T,d)d\varepsilon_i\varepsilon_0}{d_i P(T,E,d)}$, which is in agreement with the published data, at least qualitatively. Let's take $d$ and $T$ dependences as examples here. We know that $\alpha = \alpha_0 + \alpha'/d$ [19]. So $\alpha d = \alpha_0 d + \alpha'$ . When $d$ decreases $P(d)$ measured keeps relatively constant at $d$>30 nm and shows a decay rate slower than linear behavior at $d$<30 nm [9]. In this case, the model [Eq (11)] predicts $t_N$ decreases (i.e. polarization relaxes faster) as $d$ decreases, consistent with the data previously published [10]. In other words, it indicates that the fully poled single-domain state of the ultrathin Pt/PZT/Pt structure becomes *kinetically* unstable as $d$ decreases and it breaks up into multi-domain state immediately (e.g. < 1 μs, depending on the magnitude of $E_{dep}$) with $P$ approaching zero with time upon removal of the external poling field.

Since $\alpha \sim P_s^3/T$ and $P_s \sim \left(\frac{T_c - T}{T_c}\right)^{1/2}$ in mean-field theory [20], we have $\frac{\alpha(T)}{P(T)} = \frac{T_c - T}{T \cdot T_c}$ . From Eq (11), one can see that $t_N$ decreases (i.e., polarization relaxes faster) as $T$ increases, which is also in good agreement with the results in the literature on both short and long-time retention studies [3, 24]. Note that retaining its polarization at ~100 ºC for 10 years is a key requirement for a ferroelectric capacitor for commercial uses. According to Eq (11) we also note that choosing a ferroelectric material with larger $P_s$ without improving the interface properties (therefore $E_{dep}$) is not a good way to solve the problem of retention. That is because the shelf-life doesn't increase linearly with $P_s$ as some may expect; in contrast it decays exponentially with $P_s$. Indeed, the polarization relaxation curves for $P_{M_0} = 40$ and $P_{M_0} = 30$ μC/cm$^2$ converges after only ~ 1 μs (not shown). This consideration favors Pt/SBT/Pt [or $RuO_2$/PZT/$RuO_2$], believed to be "dead layer" free [25], over Pt/PZT/Pt, although SBT has lower $P_s$.

(4) This model neglects the effects of internal charge compensation (i.e., defect movement at the film interior) and possible external charge compensation (i.e., possible charge injection through the passive layer, if it exists, from the electrode) of the depolarization field, since these two factors are very sample-dependent and have many unknown parameters. However, we notice that these two factors slow down the retention loss and therefore elongate device life. Therefore, the present model actually gives the *lowest* limit for the shelf life of a model device. [Note that incorporating the charge-compensation effects into the model is not difficult if we know all the details about these processes for a given ferroelectric capacitor].

In summary, a model free of adjustable parameters has been built for retention loss at both short and long time scale in ferroelectric capacitors. We show that the predictions of this model are in good agreement with the results previously published. Particularly, it accounts for why power-law gives better fitting than linear-log relation on a short-time scale ($10^{-7}$ s to 1 s) and why a stretched exponential function provides more precise explanation than a linear-log function on a long-time scale (>100 s), as reported in the literature. Furthermore, more severe retention losses in thinner films and at higher temperatures have also been correctly predicted by the present model.

**Figure Captions:**

Fig 1 (Color online) the (*Y*, *X*) profiles of $\left(\frac{P_{M_0-N}}{P_{M_0}}=\frac{M_0-2N}{M_0}, t=\sum_{i=1}^{N} t_i\right)$ for $M_0$=100, 500 and 1000 according to Eq (10). Data points for 10 years and the age of the Earth are also labeled.

Fig 2 (Color online) the power-law and stretched exponential plots of the profile in Fig 1 with $M_0$=1000.

Fig 3 (Color online) the power-law plot and fitting for the data from $10^{-6}$ s to 1 s (filled squares), and the stretched exponential plot and fitting for the data from 100 s to $10^{6}$ s (filled triangles).

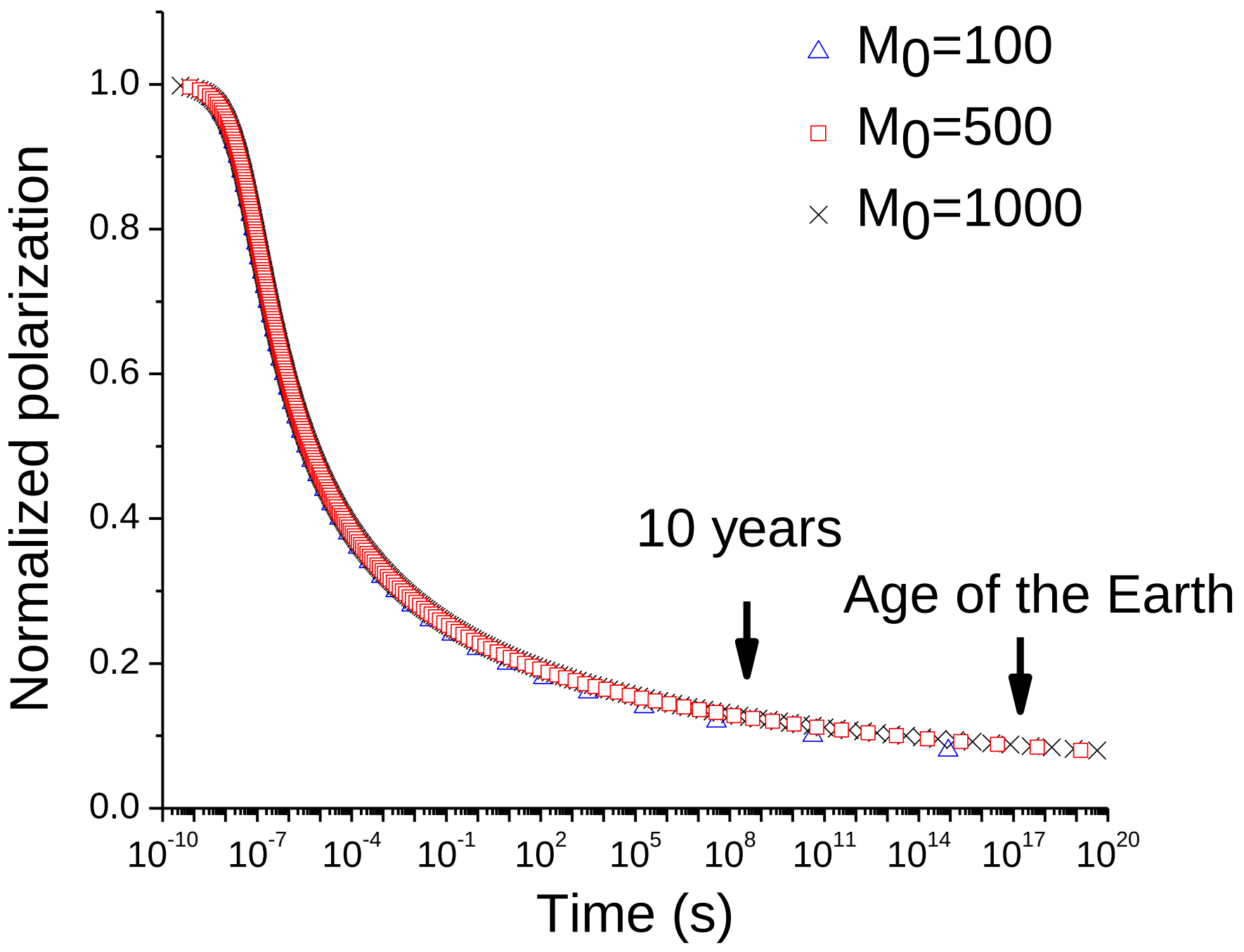

Fig 1

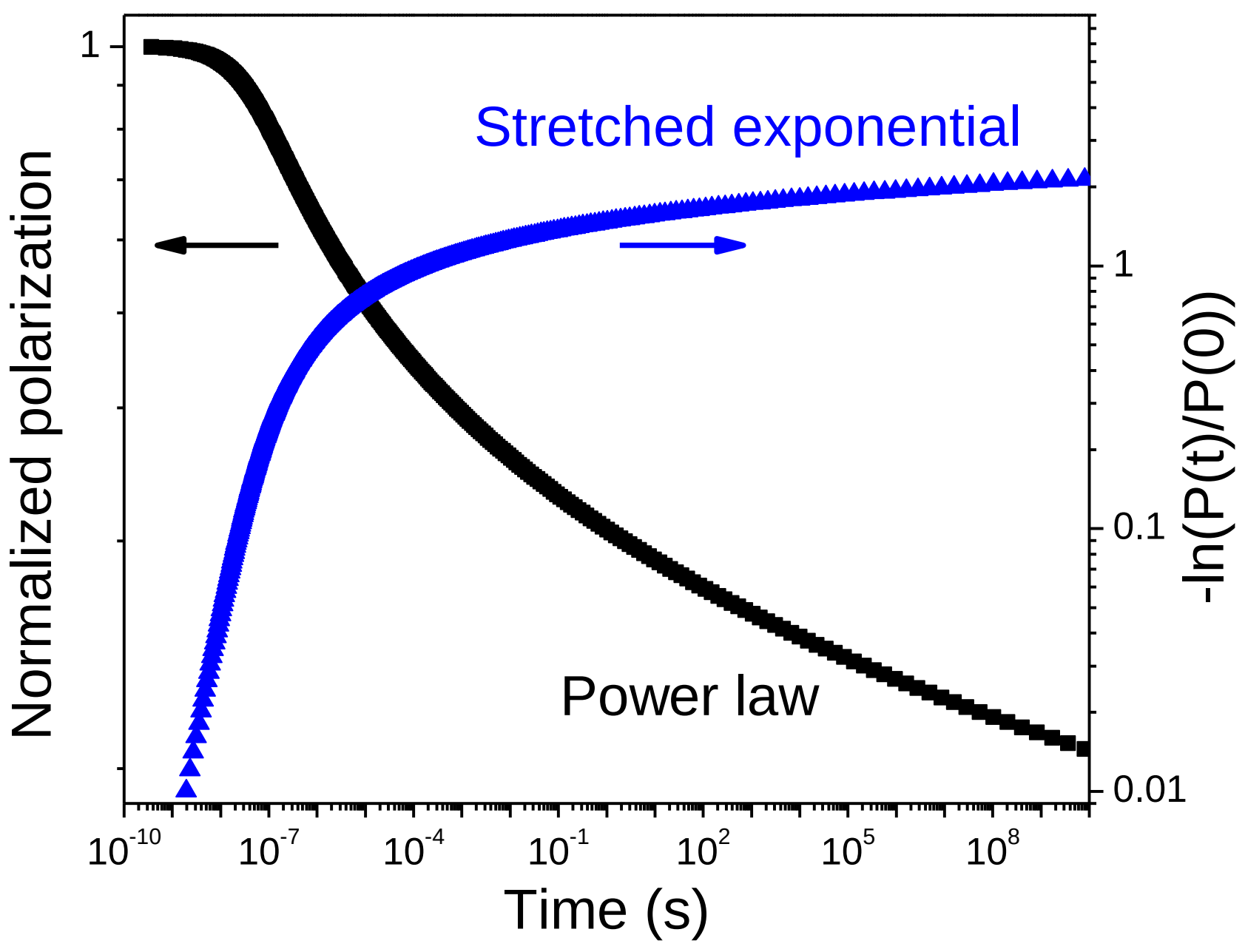

Fig 2

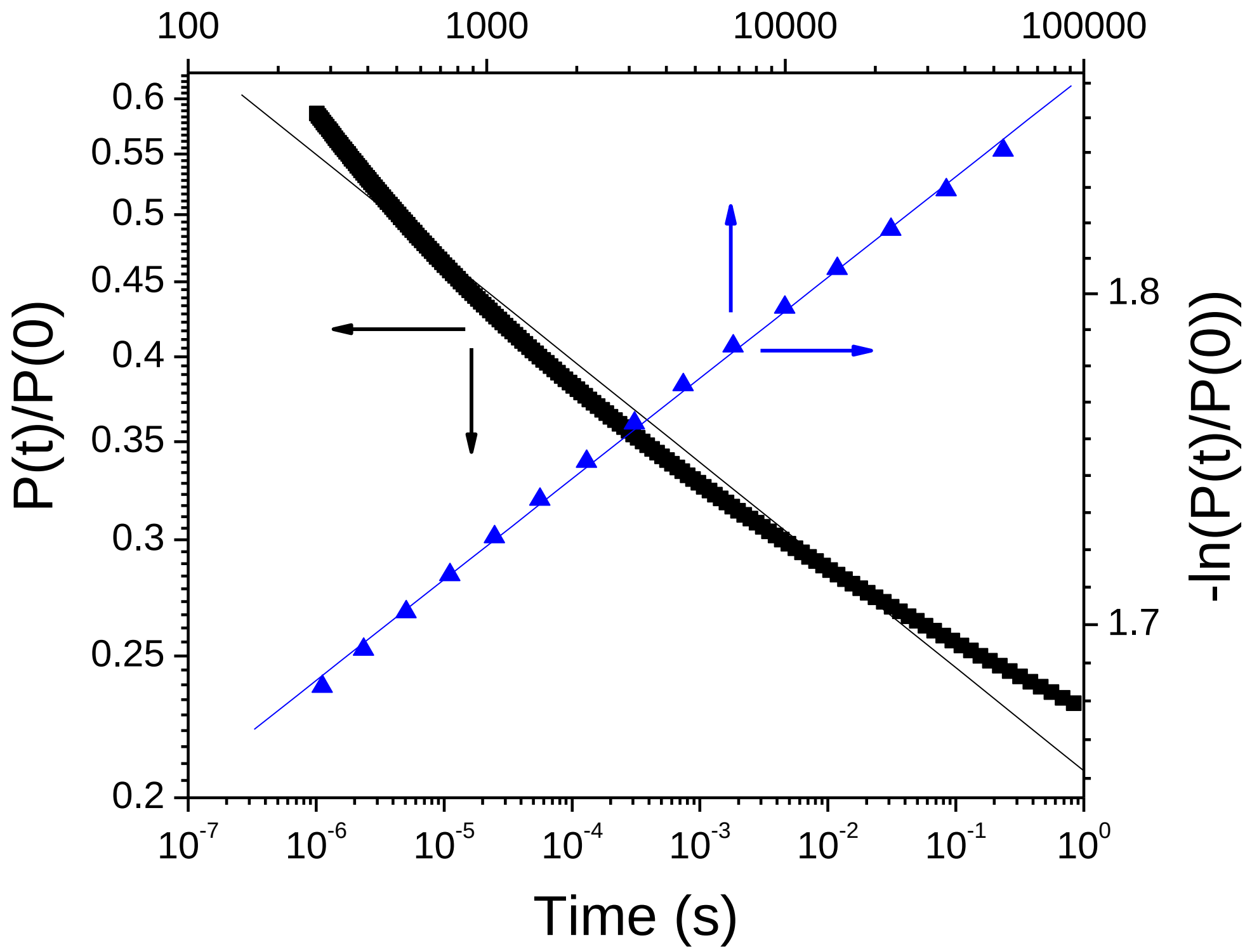

Fig 3